\documentclass[twocolumn,showpacs,preprintnumbers,amsmath,amssymb,floatfix,10pt]{revtex4}
\newcommand \beq {\begin{equation}}
\newcommand \eeq {\end{equation}}
\newcommand \ben {\begin{eqnarray}}
\newcommand \een {\end{eqnarray}}

\usepackage{wrapfig}
\usepackage{times}
\usepackage{epsfig}
\usepackage{graphicx}% Include figure files
\usepackage{color}
\usepackage{dcolumn}% Align table columns on decimal point
\usepackage{bm}% bold math
\usepackage{nomencl}
\usepackage{amsmath}
\usepackage{algorithm}
\usepackage{algorithmic}
\usepackage{setspace}

\makeglossary

\begin{document}
\bibliographystyle{plain}

\title{Competition between Surface Energy and Elastic Anisotropies in the Growth of Coherent Solid State Dendrites}

\author{Michael Greenwood, Jeffrey J. Hoyt, Nikolas Provatas}

\affiliation{ Department of Materials Science and Engineering,
McMaster University, 1280 Main Street West, Hamilton, Ontario, L8S
4L7, Canada}

\begin{abstract} A new phase field model of microstructural
evolution is presented that includes the effects of elastic strain
energy. The model's thin interface behavior is investigated by
mapping it onto a recent model developed by Echebarria et al \cite{Kar04}.
Exploiting this thin interface analysis the growth of solid state
dendrites are simulated with diffuse interfaces and the phase
field and mechanical equilibrium equations are solved
in real space on an adaptive mesh. A morphological competition between
surface energy anisotropy and elastic anisotropy is examined.  Two
dimensional simulations are reported that show that solid state dendritic
structures undergo a transition from a surface dominated [10]
growth direction to an elastically driven [11] growth direction by
changes in the elastic anisotropy, the surface anisotropy and the
supersaturation. Using the curvature and strain corrections to the
equilibrium interfacial composition and linear stability theory for
isotropic precipitates as calculated by Mullins and Sekerka, the
dominant growth morphology is predicted.
\end{abstract}

 \maketitle

\section{Introduction}

Recent years have seen the increased use of phase field modeling of free-boundary
problems related to microstructure growth.  This method has the
advantage in that it avoids explicit front tracking by making phase
boundaries spatially diffuse through the use of one or more order parameters,
$\phi$, which vary continuously across interfaces. This method of simulation
has been very successful in modeling dendritic solidification
\cite{Boe02,Pro05}.  In particular, Phase Field theory has confirmed
the predictions of the
microscopic solvability theory of dendrite growth
\cite{Kessler1988,Ben1984,Meiron1986}.  That is, the anisotropy in the
surface energy ($\gamma$), although small,
controls the growth rate and morphology of
the resultant precipitate \cite{Kar96}.

More recently the phase field approach has been used to study
dendrite growth directions in cases where multiple sources of
anisotropy can compete to control the resultant morphological
structure.  Haxhimali et al.  \cite{Haxhimali2006} showed that the
dendrite growth direction in Al-Zn alloys can change continuously
from (100) to (110) with the addition of solute. The authors
explained their observations as an interplay between the four fold
and six fold anisotropy parameters \cite{HoytRev}, was conjectured
to vary as a function  of solute.  Analogously, Provatas et al.
\cite{provatas2003} modeled the change from (100) directed
dendirites to a seaweed structure as the surface energy anisotropy
competed with the direction of the temperature gradient.

The phase field method has also been used to model the effects of
elastic and plastic effects  on microstructural evolution
\cite{Morin1995,lowengrub1997,Aguenaou1998,Wang2002A,
Wang2002B,Yeon2005,Steinbach2006}. In these studies the elastic
strain energy was added to the phase field energy functional and the
elastic modulus tensor ($C_{ijkl}$) was assumed to be a function of
the phase field $\phi$. Many of these studies have focused on the
interaction of elastic fields on growth and coarsening rates of
solid state particles  \cite{Chen2004,Zhu2004}. These studies have
shown that elastic effects play an important role in controlling
growth rates and morphology, consistent with previous analytical
predictions of Laraia et al. \cite{Laraia1988} and Voorhees et al
\cite{Voorhees1992}. Energy minimization techniques also showed that
the elastic free energy drives the interface to become unstable for
mismatched elastic coefficients under both applied and self strain
\cite{Kassner2001A,Kassner2001}. Other studies have also
incorporated the effect of mobile dislocations in phase field
modeling \cite{Haataja2001}. They have examined such effects as the
role of dislocations on the coarsening rate during spinoidal
decomposition \cite{Mah04}.

Another phenomenon of both scientific and practical interest, and
that has has not received enough theoretical study, concerns the
morphology  and growth rate of solid  state dendrites when two
sources of anisotropy can compete.  Anisotorpy can result from the
surface energy ($\gamma$) or the more interesting result of
anisotropy emerging in the elastic modulus tensor ($C_{ijkl}$).
Experimentally it has been found that dendritic structures can
appear in the solid state \cite{Malcolm67,Husain1999} provided
specific elastic and kinetic relationships between the parent and
precipitate phases are met. As discussed by Husain et al.
\cite{Husain1999}, solid state dendrites are made possible if the
precipitating phase and the parent phase have similar atomic lattice
structures, there is a high rate of atomic transfer between the two
phases and the transformation rate is controlled by diffusion in the
presence of an anisotropy.  More exotic structures can also be found
through competing mechanisms leading to interface instabilities as
shown by Yoo \cite{Yoo2005}.

This paper studies solid state dendritic growth in the presence of competitive
interactions between surface energy and elastic anisotropy. It quantifies a
transition between dominant growth morphologies controlled by each of these two
anisotropic effects. Section \ref{Model} presents the phase field model
used in this study.  Section \ref{Ele:GTC} examines the effect
of isotropic strains on the equilibrium composition. Section
\ref{AnisApprox} provides an approximation to the anisotropic strain
field at the interface of the precipitate. Section \ref{Morph:MCP}
illustrates the effect of the surface energy anisotropy, elastic energy
anisotropy and supersaturation on the growth morphology of dendritic growth.
Finally, the transition between a surface energy controlled morphology and
an elastic energy controlled morphology is quantified in Section
\ref{Characterization}.

\section{Phase Field Model\label{Model}}

This section derives a phase field free energy for solid state
transformations, which assumed a dilute alloy free energy for the
bulk chemical thermodynamics. To this is added an additional
contribution to account for  elastic  free energy. Corresponding
phase field equations are then derived from this free energy. It is
shown that the phase parameters can be related according according
to the thin interface asymptotics of Ref.~\cite{Kar04} to a good
approximation. The thin interface limit allows for significant
efficiency of simulations of the phase field equations, particularly
when combined with novel adaptive mesh refinement algorithms
\cite{Pro99,Badri2007}.

The bulk chemical free energy is given by
%%%%%%%%%%%%%%%%%
\ben G(\phi,C,\epsilon_{ij})=f(T_m)-S(\phi)(T-T_m)\nonumber
\\+\frac{RT}{\nu} \left( c \ln c -c\right) +E(\phi) c +
f_{el}(\phi)\label{EE:FreeEnergy} \een
%%%%%%%%%%%
$C$, $\phi$ and $\epsilon_{ij}$ are the solute concentration, order parameter and
total strain fields, respectively. The parameter $R$ is the gas
constant, $\nu$ is the molar volume, $T$ is the temperature and
$T_m$ i the melting temperature of component $A$
of a dilute binary alloy.
%%%%%%%%%%%%%%%%%%%%%%%%%%%%%%%
The elastic free energy is defined by Hooke's law, given by
\begin{equation}
f_{el}(\phi) = C_{ijkl}(\phi)(\epsilon_{ij} -
\epsilon_{ij}^*(\phi))(\epsilon_{kl} - \epsilon_{kl}^*(\phi))
\label{EE:fel}
\end{equation}
where $\epsilon_{ij}^*(\phi)$ is the phase dependent eigenstrain, given by
$\left(\epsilon^*_{ij}(\phi)= \frac{1+g(\phi)}{2}\epsilon^{*A}_{ij}
+ \frac{1-g(\phi)}{2}\epsilon^{*B}_{ij}\right)$. Here
$\epsilon^{*A,B}_{ij} = \epsilon^*\delta_{ij}$ is the hydrostatic
lattice eigenstrain with $\epsilon^* = \frac{a_A-a_B}{a_A}$ and
$a_i$ is the lattice parameters in each phase. The phase dependent
elastic tensor is interpolated across the interface
 by $C_{ijkl} = \frac{1+g(\phi)}{2}C_{ijkl}^{A} +
\frac{1-g(\phi)}{2}C_{ijkl}^{B}$. A convenient choice for the
interpolation function $g(\phi)$ which maintains the bulk phases at
$\phi = \pm 1$ is given as \ben g(\phi) = \frac{15}{8}\left(\phi -
\frac{2}{3}\phi^3+\frac{1}{5}\phi^5\right)\een

The elastic energy can be re-cast as a function of $g(\phi)$,
leaving the elastic free energy in the form of
\begin{equation}
f_{el} ={\it Z_3}  \left( g(\phi)  \right) ^{3}+{\it Z_2}  \left( g
 \left( \phi \right)  \right) ^{2}+{\it Z_1} g(\phi) +{ \it Z_0}
 \label{EE:felNV}
\end{equation}
Each pre-factor in the polynomial of $g(\phi)$ (i.e., $Z_3$, $Z_2$, $Z_1$ and $Z_0$)
is a function  dependent on the elastic modulus tensor. Explicit
expressions for these prefactors are given in the appendix (Appendix
\ref{Ele:Cubic}) for the cubic elastic modulus tensor in
two-dimensions.

It is straightforward to show using the common tangent technique that the
equilibrium composition, denoted $C_b^{eq}$, is modified by elasticity according
to,
\begin{equation}
{\it C_b}= C_b^{eq} - G_{ele}= {\frac { \left( T-{\it T_m} \right)
L\nu}{{{\it T_m}}T R
 \left( 1-k \right) }}-\frac{2\nu}{TR}{\frac { \left( {\it Z3}+{\it Z1}
 \right) }{ \left( 1-k \right) }}
 \label{Ele:Cb}
\end{equation}
 where ${\it
C_b^{eq}}={\frac { \left( T-{\it Tm} \right) L\nu}{{{\it T_m}}T R
 \left( 1-k \right) }}$ is the equilibrium coexistence line corresponding to
 the parent phase
  in the absence of elasticity,  $G_{ele} = \frac{2\nu}{TR}{\frac { \left( {\it
Z3}+{\it Z1} \right) }{ \left( 1-k \right) }}$ is the correction to
 the phase diagram due to a local change in the strain and
 $k= \frac{C_A^{eq}}{C_B^{eq}}$ is the partition coefficient.

\subsection{A Phase Field Model For Elastically Influenced Phase
Transformations \label{Ele:PFM}}

By applying the usual dissipative dynamics for the order parameter,
mass conservation for concentration and strain relaxation for
the strains, the following kinetic equations for the order parameter
$\phi$, undercooling $U \equiv \frac{e^u-1}{1-k}$ (where $k$ is the partition
coefficient and $u$ is defined in below) and the strain fields $\epsilon_{ij}$
are derived:\\
Phase Mobility: \ben \tau A(\hat{n})^{2} \frac {\partial \phi}
{\partial t} =
\hat{\nabla} \cdot [W^{2}A(\hat{n})^{2} \hat{\nabla} \phi]  \nonumber \\
+ \hat{\nabla} \cdot \Big ( |\hat{\nabla} \phi|^{2} W^{2} A(\hat{n})
\frac {\partial A(\hat{n})} {\partial (\hat{\nabla} \phi)} \Big )\nonumber\\
 -( \phi^{3} - \phi) - \lambda (1+B)  U (1-\phi^{2})^{2}\label{EE:PDM}
 \een
Chemical Diffusion: \ben
 \Psi \frac{\partial U}{\partial t} =
\vec{\nabla}\cdot(D
\tilde{q}(\phi) C_b\vec{\nabla}U) \nonumber \\
+ (1+ (1-k)U) C_b\left(1+\frac{\alpha Z_2}{C_b
(1-k)}\right)\frac{\partial
\phi}{\partial t} \nonumber \\
- (1+ (1-k)U)\left(\frac{k+1}{1-k}-\phi\right)\frac{\partial
C_b}{\partial t} \label{EE:chemdiff} \een Strain relaxation:
\ben\frac{\partial \sigma_{ij}}{\partial x_j}
\equiv\frac{\partial}{\partial x_j} \frac{\delta
G(\phi,C,\epsilon_{ij})}{\delta \epsilon_{ij}^{el}}
=0\label{Ele:StEle}\een
The explicit form of $B$ in equation
\ref{EE:PDM} is \ben B =\left(
 \frac{\alpha
Z_1 }{(1-k)C_b^{eq}}-1\right)\frac{\alpha
Z_2}{2C_b(1-k)}g(\phi)\nonumber \\
 -\frac{\alpha Z_1}{2C_b^{eq}(1-k)}\label{EE:B}\een
where $C_b$ is the equilibrium composition corrected for the strain
by equation \ref{Ele:Cb}, i.e., $C_b = C_b^{eq}-\frac{1}{2}\frac{\alpha Z_1}{(1-k)}
 \label{EE:PECb}$ with $C_b^{eq}$ being the equilibrium composition
 in the absence of elasticity  $\alpha=\frac{4\nu}{RT}$  ($Z_3$ has been neglected here, as
discussed below). The quantities $W, \tau$ and $\lambda$ are the
usual phase field  parameters setting the dimensions of space and
time \cite{Kar96,Kar04}. The function $A(\hat{n}) \equiv
1+\epsilon_4cos(4\theta)$ controls the four-fold surface energy
anisotropy, where $\theta$ is the angle of orientation between the
interface normal $\hat{n}$ and a reference axis.

The dimensionless undercooling $\left(U \equiv \frac{e^u-1}{1-k}\right)$
in equation~\ref{EE:chemdiff} is modified from its form in Ref.~\cite{Kar04}
to incorporate an elastic correction to the equilibrium composition.
This results in \ben e^u =
\frac{2C}{C_b}\left(k+1 -(1-k)g(\phi)-\frac{\alpha Z_2}{
2C_b}(1-g(\phi)^2)\right)^{-1} \label{EE:eu}\een where $Z_2$ is the
prefactor to $g(\phi)^2$ in the elastic free energy. The function
$\Psi$ modulates the diffusion through the
interface correcting for the diffuse nature of $\phi$ and is given
by
 \ben \Psi = C_b(k+1 -(1-k)\phi-\frac{Z_2
\alpha}{2 C_b}(1-\phi^2))\label{Ele:Psi}\een

Two-sided diffusion is controlled by the function $\tilde{q}(\phi) =
q(\phi)\frac{\Psi}{C_b}$ where $q(\phi)$ modulates the diffusion in
the two phases, to simulate equal diffusion coefficients $q(\phi) =
1$.  The dimensional diffusion coefficient is denoted $D$.The
value of elastic coefficients used herein is a actually
reported as $\alpha C_{ijkl}\label{EE:dimCijkl}$.

The phase field equations above  are derived in the limit where
$Z_3 \rightarrow 0$, the condition for a small difference in the
elastic coefficients in either phase. This condition also
holds for all materials in which $\frac{Z_3}{Z_1} << 1$ holds.  Generally
speaking $Z_1 > Z_3$ even in the most extreme disparities of the elastic
coefficients, so this assumption is not unreasonable.

Following the procedure for the derivation of the phase field model
presented in \cite{Kar04}, the constants $W_o$, $\tau_o$ and
$\lambda$ are inter-related by the asymptotic analysis of Ref.
\cite{Kar96}. This analysis maps the phase field model above,
without elasticity, onto the sharp interface limit governed by the
Gibbs-Thomson condition $C_{int}  = C_{eq} - \Delta C
d_o(\vec{n})\kappa - \Delta C \beta(\vec{n})V$, where $d_o$ is the
capillary length, $\beta_k(\vec{n})$ is interface kinetic
coefficient and $\Delta C$ is the concentration jump across the to
phase interface. Echebarria et al. showed that for the case of
vanishing kinetic coefficient (appropriate for the study of solid
state denderites) the following relations must be obeyed:
$\frac{d_o}{W}=0.8839 \lambda$ and $D\equiv \bar{D}\tau/W^2=0.6267
\lambda$ where $\bar{D}$ is the diffusion coefficient in dimensional
units ($m^2/s$). These relationships arise by an expansion of the
phase ($\phi$) and composition ($C$) inside the interface, which is
matched to solutions outside the interface such as to emulate the
sharp interface boundary conditions.

It expected that for the phase field model presented in
this work the thin interface  relationships between $W_o$, $\tau_o$ and
$\lambda$ will be, to lowest order, the same as those for the
model of Echebarria et al. based on two observations. The first is drawn from the work of
Yeon et al. \cite{Dong2005}, who showed that, to first order, the inclusion of elasticity
is decoupled from the curvature and kinetic effects in the
Gibbs-Thomson condition, which
reads  $C_{int}  = C_{eq}- \Delta C d_o(\vec{n})\kappa -
\Delta C \beta_k(\vec{n})V - G_{ele}$.  In their asymptotic matching procedure
the derivation of $d_o$ and $\beta_k(\vec{n})$ remained unaffected
by the presence of the $G_{ele}$ contribution.

The second observations validating the use of the thin interface analysis of
Ref. \cite{Kar04} for the phase field model presented here is that for our
simulations it was foind that $\frac{\partial C_b}{\partial t} << 1$ in Equation
\ref{EE:chemdiff} and $B\ll1$ in equation~\ref{EE:PDM}. The first
quantity was found numerically to be at least an order
of magnitude smaller than the other terms in its equation.  It was likewise
also found that  $B << 1$ (which includes $Z_2\ll1$).   Both $B$ and
$\frac{\partial C_b}{\partial t}$ are zero in
the limit of zero strain, but are also small for small lattice
eigenstrains.  For larger strains these terms are still valid as long as the
interfacial velocity ($V$) is small. In effect a small velocity
eliminates any excess kinetic effects that arise at the interface.
In this work only small eigenstrains ($\epsilon^* \approx 0.005$)
and small growth rates ($V < 1\mu m/s$) are considered.

It should be noted that the small variable $B$
accounts for the effect of a variable strain field on the
development of the precipitate. It plays an analogous
role to the temperature correction of $\lambda$ that is used
when simulating solidification of alloys with \emph{non-linear}
coexistence phase boundaries \cite{Tong2007}.

\section{Growth of an Isotropic Second Phase Precipitate with
Coherent Interfaces in an Isotropic Parent Phase \label{Ele:GTC}}

This section examines the interface compositions of an isolated second phase
precipitated into a parent phase, which
is grown under isotropic conditions in both the parent and precipitate
phases. The surface energy is made isotropic by setting the surface
energy anisotropy coefficient $\epsilon_4=0$. The elasticity equations are
formulated in terms of cubic tensor coefficients, ie
$C_{11}$,$C_{12}$ and $C_{44}$ as described in appendix
\ref{Ele:Cubic}. For isotropic linear elastic coefficients the cubic
elastic terms are related by, \ben C_{44} =
\frac{1}{2}(C_{11}-C_{12}) \label{Ele:IsoC44}\een
\begin{figure}[h]
    \centerline{\includegraphics*[height=2in,width=3in]{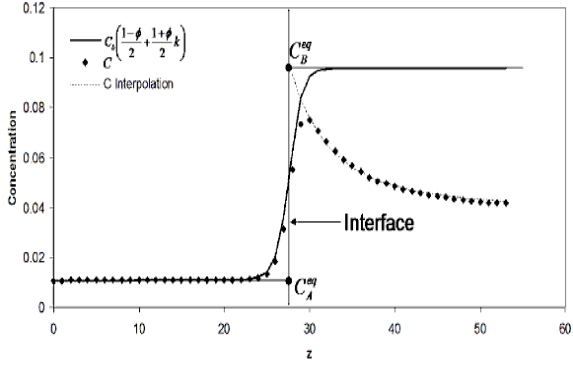}}
    \caption[]{\linespread{0.75}\em{A circular isotropic coherent precipitate is grown in an alloy with
    initial composition of $C_o=0.04$ with a hydrostatic lattice eigenstain of $\epsilon^*=0.005$. A cross section of
    the composition is shown with the corresponding equilibrium composition corrected for strain,
    $C_b\left(\frac{1-\phi}{2}+\frac{1+\phi}{2}k\right)$.  The compositions are interpolated to the center of the interface
    and are found to be in good agreement with the sharp interface boundary condition.}} \label{Ele:GTInterC}
\end{figure}
where for this simulation $C_{11}=1011$ and $C_{12}=729$ ($C_{ij}$
is dimensionalized by $\alpha$ )
 and
the coherent hydrostatic eigenstrain is set to $\epsilon^*=0.005$.
The convergence constant of the phase field equations is $\lambda = 3$. The
equilibrium composition is $C_b^{eq} = 0.1$ with an initial alloy composition
of $C_o=0.04$ and the solute partition coefficient is set to a value of $k=0.1$. The
diffusion coefficients are set equal in both phases thus removing
the phase dependency in the diffusion coefficient (ie. $q(\phi) =1$)
in equation \ref{EE:chemdiff}.

The total domain size simulated  is $6400W$ on a side with periodic boundary
conditions. The phase field and diffusion equations are solved using
an explicit time stepping algorithm on an adaptive mesh with a grid
spacing of $dx=0.391$ at the lowest level of refinement and a time
step of $dt=0.01$.  The displacement field is solved by direct
Gauss-Seidel iteration, which is found to require
($O(N)$) operations per time step  on an adaptive mesh \cite{MikeThesis2008}.

The precipitate particle is grown and a cross section of the
composition field solved by Equation \ref{EE:chemdiff} is shown in
Figure \ref{Ele:GTInterC}. This figure indicates the corresponding value of the
equilibrium interfacial composition as calculated by equation
\ref{Ele:Cb}. $C_b$ is plotted as a function of the phase by
interpolating it through the interface to its corresponding
precipitate side value by
$C_b\left(\frac{1-\phi}{2}+\frac{1+\phi}{2}k\right)$.  The
composition field is interpolated to the interface described by the
point where $\phi=0$ both from inside the precipitate bulk and from
outside the precipitate in the parent phase. The points are denoted
$C_A^{eq}$ and $C_B^{eq}$ and are found to have excellent agreement with
the interpolations to the center of the interface.

\section{Approximation of the Strain Field Around a Precipitate
with Cubic Elastic Coefficients\label{AnisApprox}}

In this section the elastic field is analyzed around a circular precipitate
where the cubic elastic coefficients are equal in both
the precipitate and matrix phases.  The
anisotropy is entered into the cubic elastic coefficients by
introducing a deviation from the isotropic relation in equation
\ref{Ele:IsoC44} as defined here by, \ben C_{44} = \frac{1}{2}
(C_{11}-C_{12}) + \beta\label{Ele:AnisC44}\een where $\beta$ is the
deviation from the isotropic elastic coefficients. $C_{11}$ and
$C_{12}$ remain unchanged. While the analytical solution to the
isotropic strain field has been derived for elliptical inclusions
under a hydrostatic eigenstrain \cite{Eshelby1957}, a solution to an
anisotropic precipitate under the same conditions is mathematically
cumbersome do deal with \cite{Ru2003} and is instead solved
here numerically.

\begin{figure}[h]
    \centering
    \includegraphics[ height=0.275\textheight]{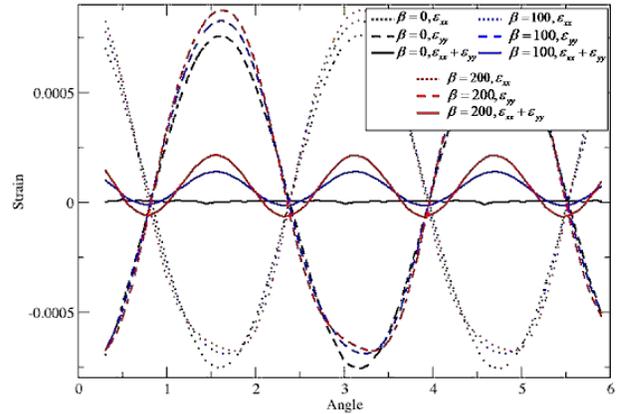}
    \caption[]{ (Colour Online) The strain fields $\epsilon_{xx}$, $\epsilon_{yy}$ and
     their sum are plotted for two particle radii with $\beta = 200$ in units of the
     model. The effect of the deviation from isotropy ($\beta = 0,100,200$) on the
    trace of the strain tensor ($\epsilon_{xx} + \epsilon_{yy}$) is illustrated by the
    solid lines.} \label{AnisElast}
\end{figure}

A circular precipitate in a parent phase with identical elastic
coefficients in both phases is considered.  The dimensionless
elastic coefficients are set to values of $C_{11}=1011$ and $C_{12}
= 729$ ($C_{ij}$ is dimensionalized by $\alpha$, ie. $\alpha
C_{ij}$), the hydrostatic eigenstrain is set to $\epsilon^* = 0.005$
and the concentration field is made constant (The elasticity here is
not influenced by compositional effects and therefore any
concentration field will produce similar results). The deviation
from elastic isotropy is studied by two controls, the particle
radius and the strength of the deviation from isotropic elasticity
($\beta$).

\begin{figure}[h]
    \centerline{\includegraphics*[height=2.5in,width=3.5in]{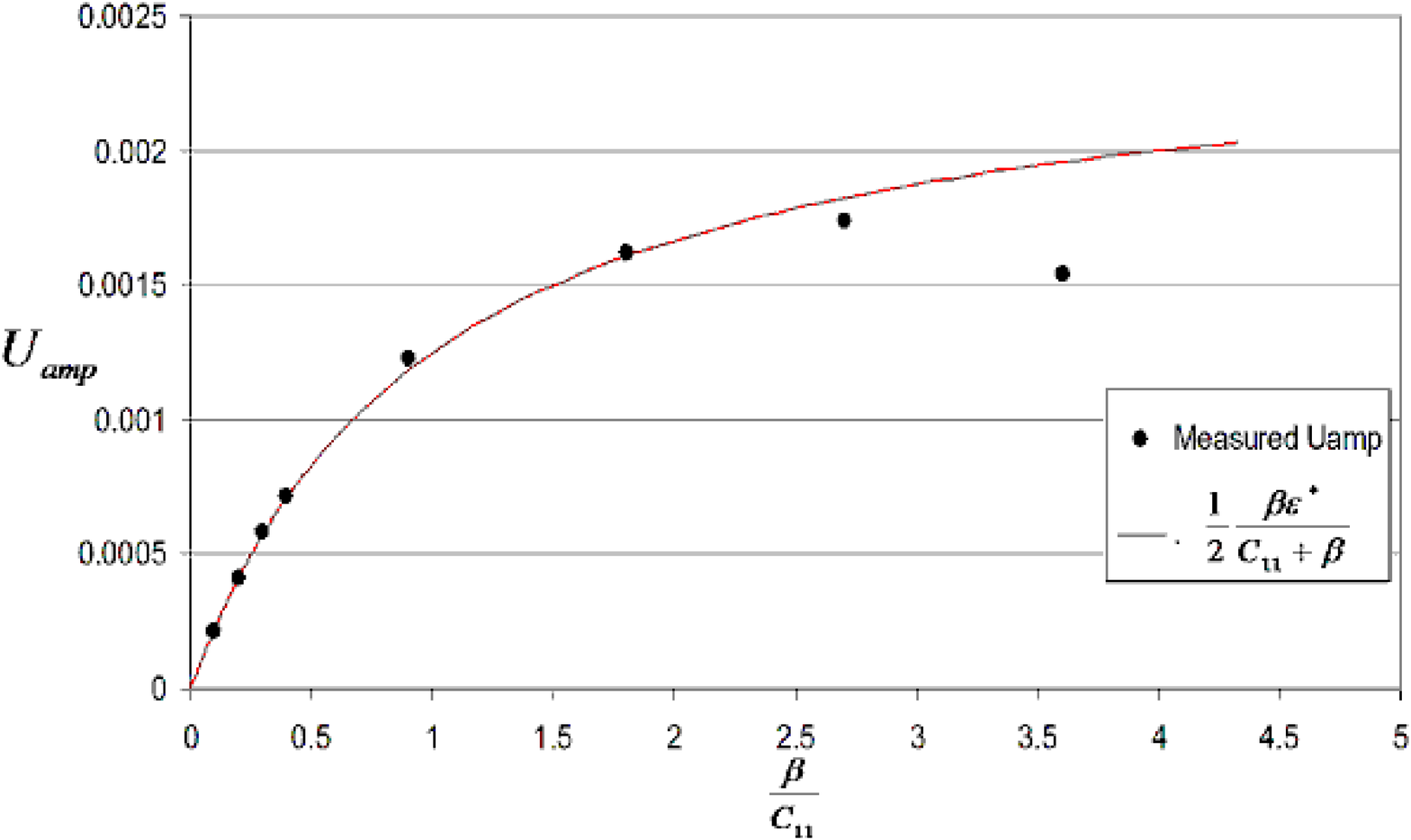}}
    \caption[]{\linespread{0.75}\em{(Colour Online) Measured values of $U_{AMP}$ plotted vs $\frac{\beta}{C_{11}}$ and the fitted prediction of $\frac{1}{2}\frac{\beta \epsilon^*}{C_{11}+\beta}$.}}
\label{AnisElast:E5}
\end{figure}

  The behaviour of the strain trace due to changes in
the strength of the elastic anisotropy ($\beta$) is studied by
holding the precipitate radius constant while increasing the value
of $\beta$. The amplitude of the trace of the strain tensor
($\epsilon_{xx}+\epsilon_{yy})$ is found to have a
strong sensitivity to variations in $\beta$.  This is illustrated in
Figure \ref{AnisElast} for $\beta = 0, 100, 200$,
where the individual strain components and their sum at the
precipitate/matrix interface are plotted as a function of the
angle $\theta$,
with zero representing the (10) direction.
Notice the four fold symmetry of the trace of the strain.
Although the strain energy depends strongly on the value of
$\beta$,
the precipitate radius is found to have
little effect on the amplitude of any perturbation to the strain
trace.

The $\theta$ dependence of the strain trace can be approximated to
lowest order by a single Fourier mode defined by,
\ben\epsilon_{xx}+\epsilon_{yy} = U_{AMP}
cos(4\theta)\label{EE:Exxyy}\een To measure the functional form for
the amplitude of the strain trace ($U_{AMP}$), the anisotropic
strength ($\beta$) is varied and the amplitude of the strain trace
($U_{AMP}$) is measured for the corresponding waveform at the
interface of the precipitate. These measured values are fitted to a
functional form, given by the equation,
\begin{equation}
U_{AMP} = \frac{1}{2}\frac{\beta \epsilon^*}{C_{11}+\beta}
\label{AnisElast:Uamp}
\end{equation}
The functional form of equation \ref{AnisElast:Uamp} shows good
agreement for all values of $\beta$ in the regime where
$\frac{\beta}{C_{11}} < 2$ as shown in figure \ref{AnisElast:E5}.
This range of anisotropies is well within the limits of the relative
anisotropic strengths that are studied here.

\section{Conditions Influencing the Morphology of Precipitates\label{Morph:MCP}}

This section characterizes three distinct controlling influences on the selection
of a dominant morphology of precipitated dendrites. These
are the anisotropies of the  elastic tensor and surface energy and the
supersaturation. The effect of each of these parameters are
systematically tested by increasing the strength of each parameter
while holding the other parameters constant.

In the following phase field simulations, $\lambda=3$ and, as
required by the sharp interface analysis \cite{Kar04}, the
dimensionless diffusion coefficient is set to $D= 0.6237 \lambda$.
The temperature is set such that the equilibrium composition is
$C_b^{eq} = 0.1$   and the partition coefficient used is $k=0.1$.
The elastic coefficients were converted to units of the model by the
elastic modulating factor $\alpha = 6.005 \cdot 10^{-9}
\frac{m^3}{J}$ and in these units the elastic coefficients are set
to a value of $C_{11} = 1011$ and $C_{12}=729$, where $C_{ij} =
\alpha C_{ij}$.

Precipitate structures are grown in a system with periodic
boundaries, where the system size is set to $6400W x 6400W$ ($W$
being the interface width, determined along with $\tau$ from the
asymptotic analysis used). The precipitates grew to sizes of at most
$2000W$ and the solution to the displacement field drops off as
$1/R$. This justifies the claim that purely isolated precipitates
are studied while using the periodic boundaries. The diffusion
coefficients and elastic coefficients have no phase dependence.  The
grid spacing is set to $dx = 0.391W$ and the explicit time step is
set to $dt = 0.01\tau$.

\subsection{Elastic Anisotropy ($\beta$)\label{Morph:beta}}

\begin{figure}[h]
    \centering
    \begin{tabular}{ccc}
        \includegraphics[height=0.2\textheight]{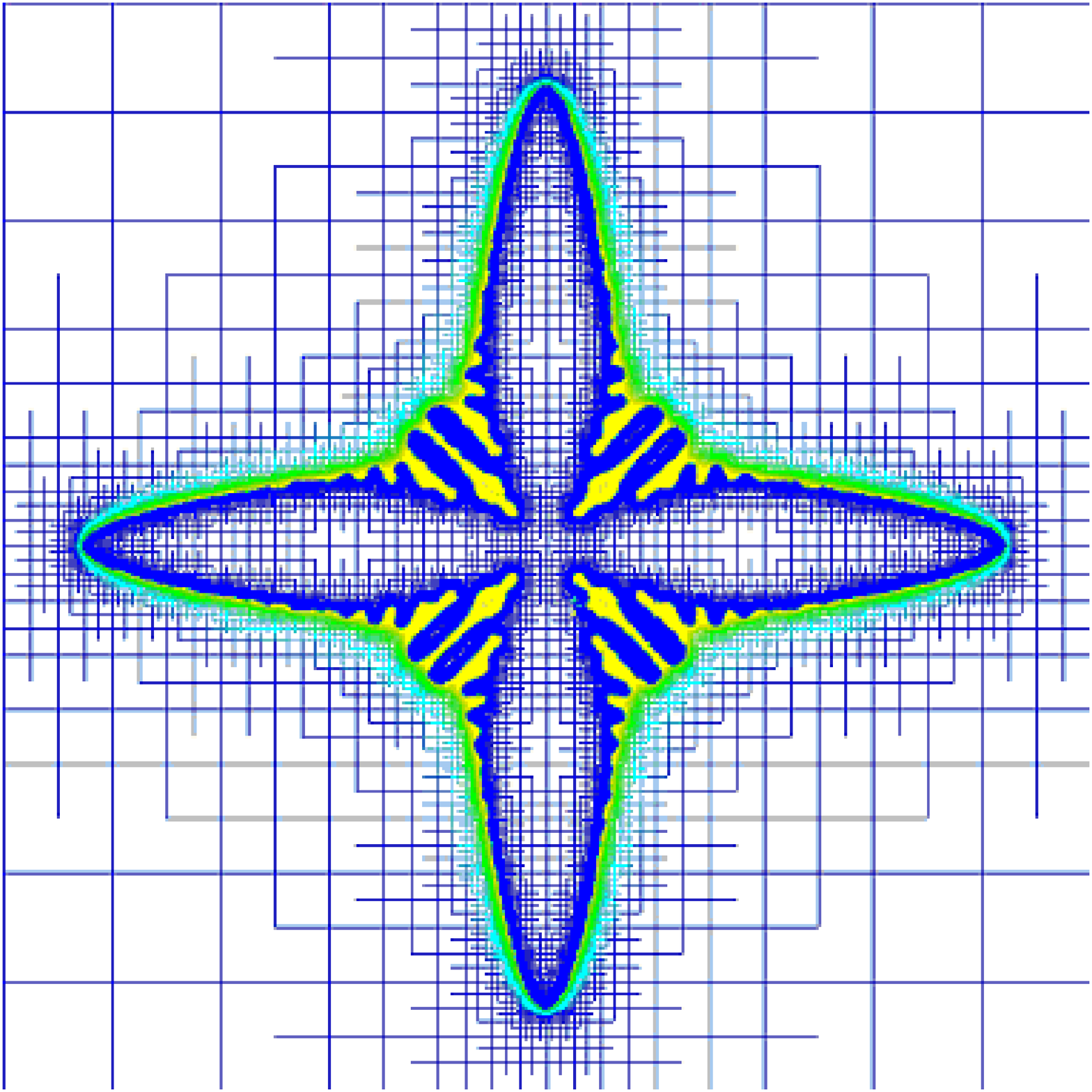}
        \\
        (a) $C_o= 0.04$,$\beta = 25$ and $\epsilon_4 = 0.01$
        \\ \includegraphics[height=0.2\textheight]{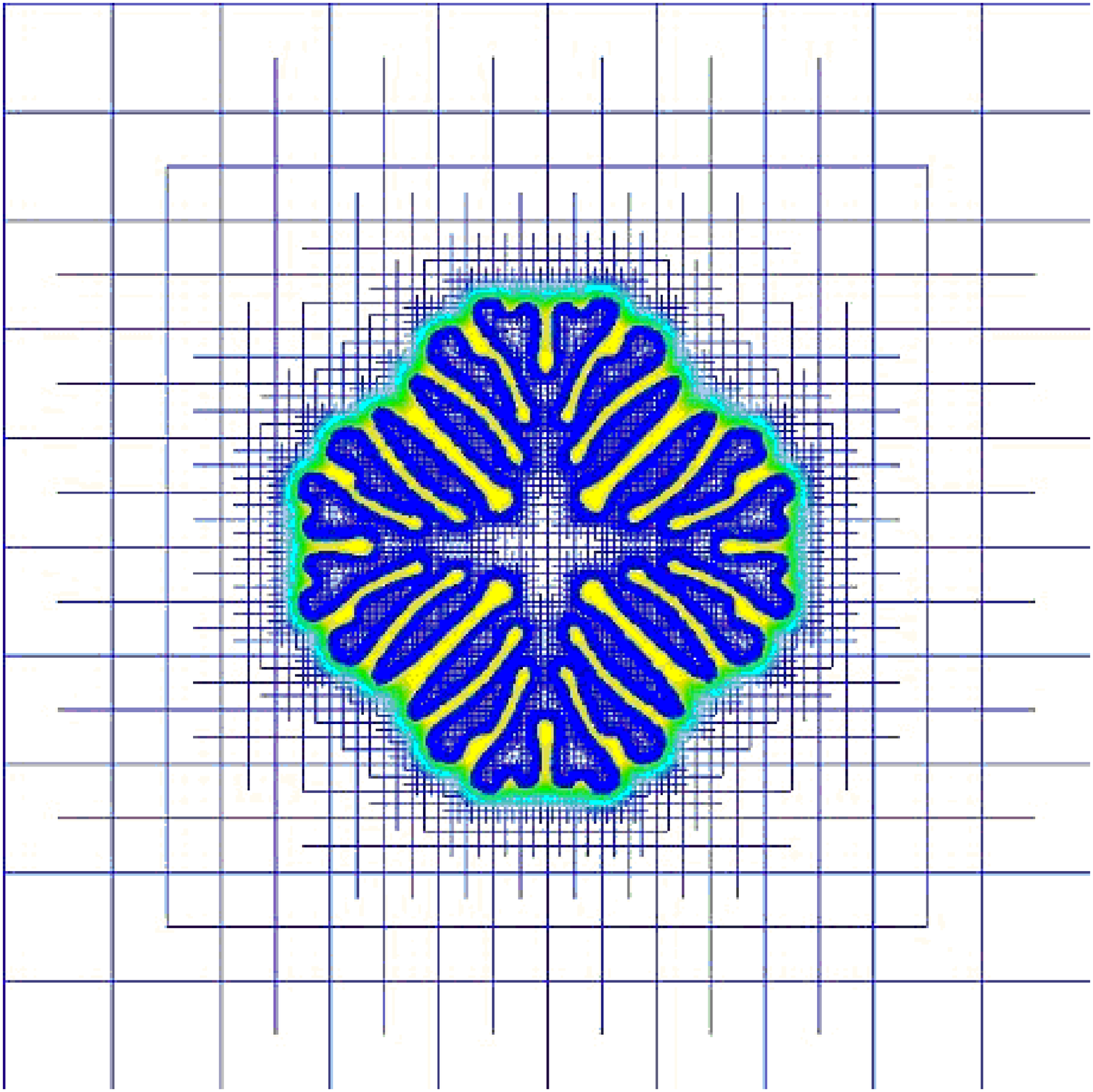}
        \\
        (b) $C_o=0.04$,$\beta = 100$ and $\epsilon_4 = 0.01$
        \\ \includegraphics[height=0.2\textheight]{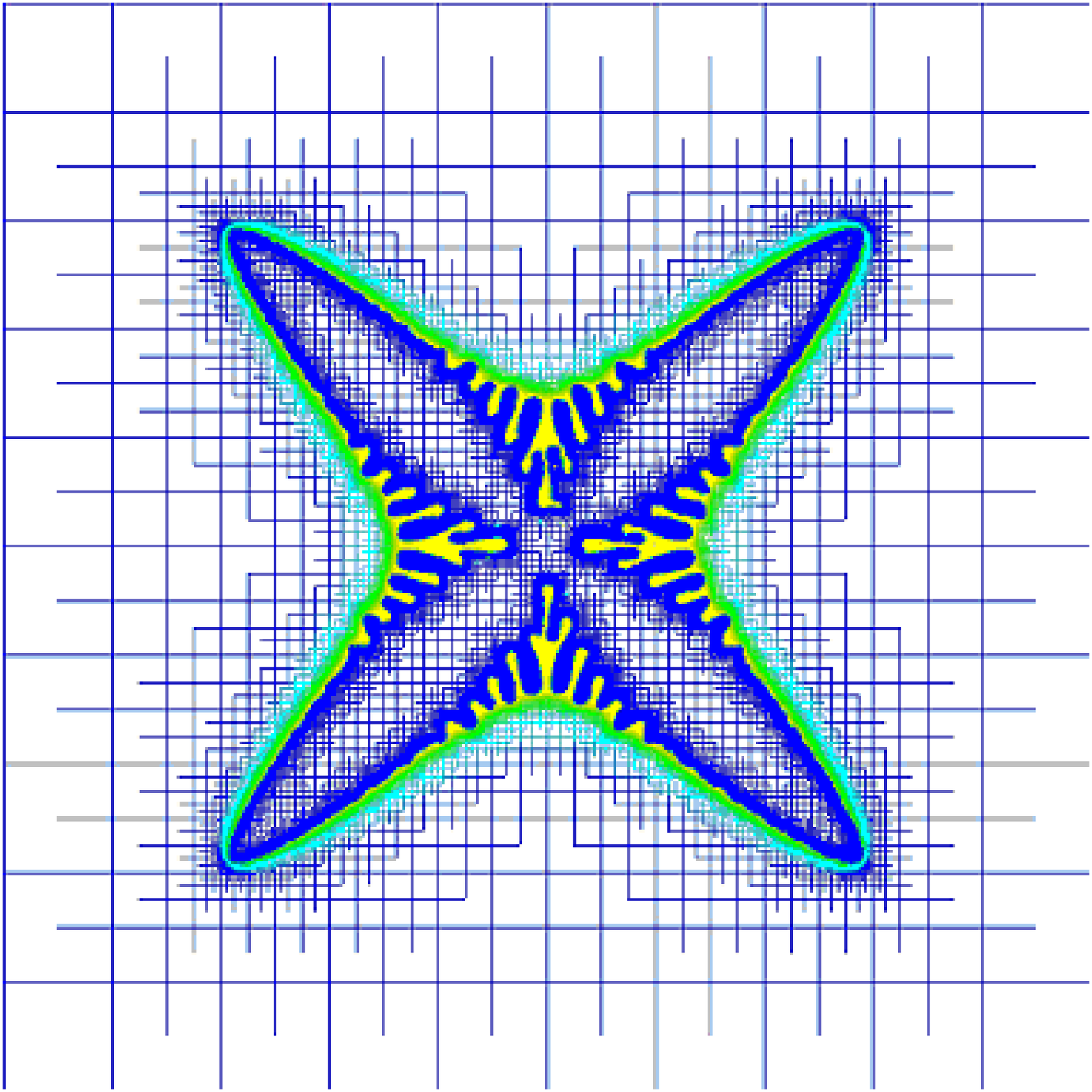}
        \\
        (c) $C_o= 0.04$,$\beta = 400$ and $\epsilon_4 = 0.01$

   \end{tabular}
    \caption[]{\emph{(Colour Online) Morphologies of growth precipitates vs.  variation in the
    elastic anisotropy in the modulus tensor through an alteration in $\beta$.
    The composition field is plotted on the adaptive mesh, blue being low concentration and yellow is a higher concentration.
    The total system size is $6400W$, plotted is an area of $2000W$ on one side. .} }\label{MorphTrans2}
\end{figure}

The elastic anisotropy emerges from the elastic tensor ($C_{11}$,
$C_{12}$ and $C_{44} = \frac{1}{2}(C_{11} - C_{12}) + \beta$). The
anisotropy of the tensor is varied by holding $C_{11}$, $C_{12}$
constant and varying $C_{44}$ through changes in $\beta$. Figure
\ref{MorphTrans2} (a-c) shows the effect of increasing the strength
of the elastic anisotropy by increasing $C_{44}$ while holding surface energy
anisotropy and supersaturation (controlled via the average composition $C_o$)
constant. From top to bottom the values of $\beta$
used are $\beta = 25,100,400$ respectively.  As can be seen in this
figure, a small elastic anisotropy causes the surface energy to
dominate and the dendrite grows in the $[10]$ direction ( Figure
\ref{MorphTrans2} (a) ). When the elastic anisotropy is increased to
sufficient strength ( Figure \ref{MorphTrans2} (c) ) the dendrite
grows in the $[11]$ direction. When the anisotropies effectively
destructively interact the resultant structure leads to an almost
isotropic growth morphology ( Figure \ref{MorphTrans2} (b) ).

\subsection{Surface Energy Anisotropy ($\epsilon_4$)}

\begin{figure}[h]
    \centering
    \begin{tabular}{ccc}
       \includegraphics[height=0.2\textheight]{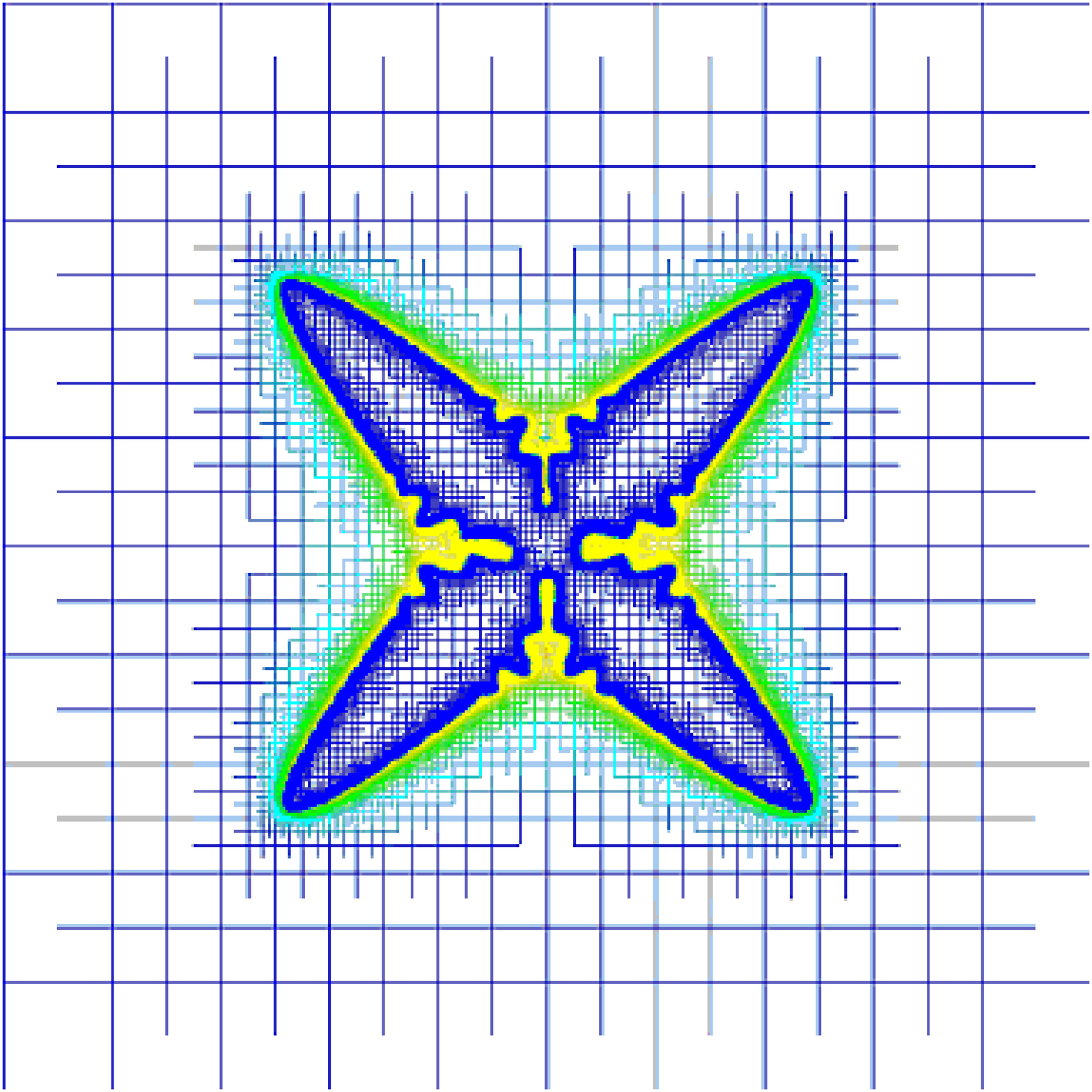}\\
       (a) $C_o=0.05$,$\beta = 400$ and $\epsilon_4 = 0.01$
       \\
       \includegraphics[height=0.2\textheight]{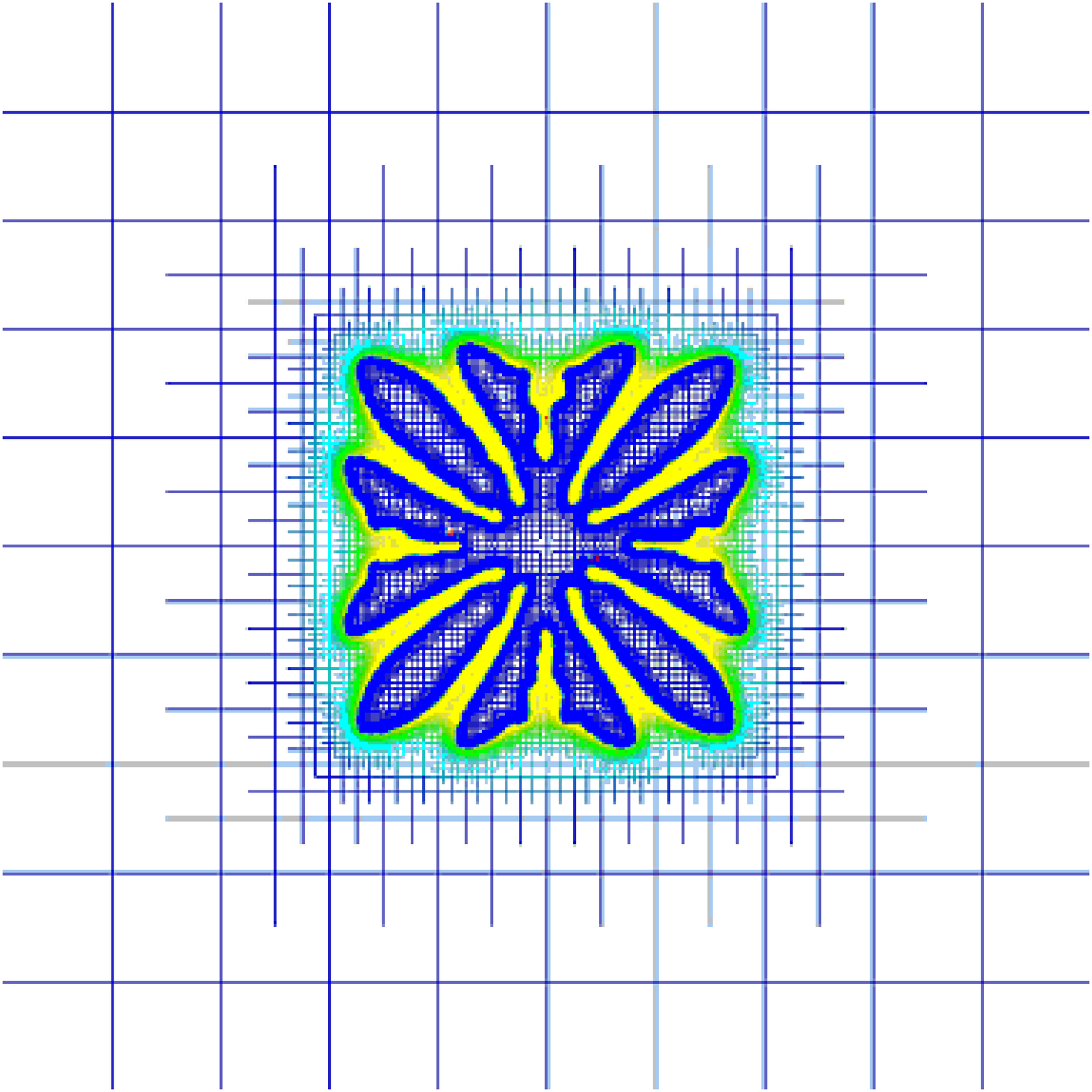}\\
       (b) $C_o=0.05$,$\beta = 400$ and $\epsilon_4 = 0.03$
       \\
       \includegraphics[height=0.2\textheight]{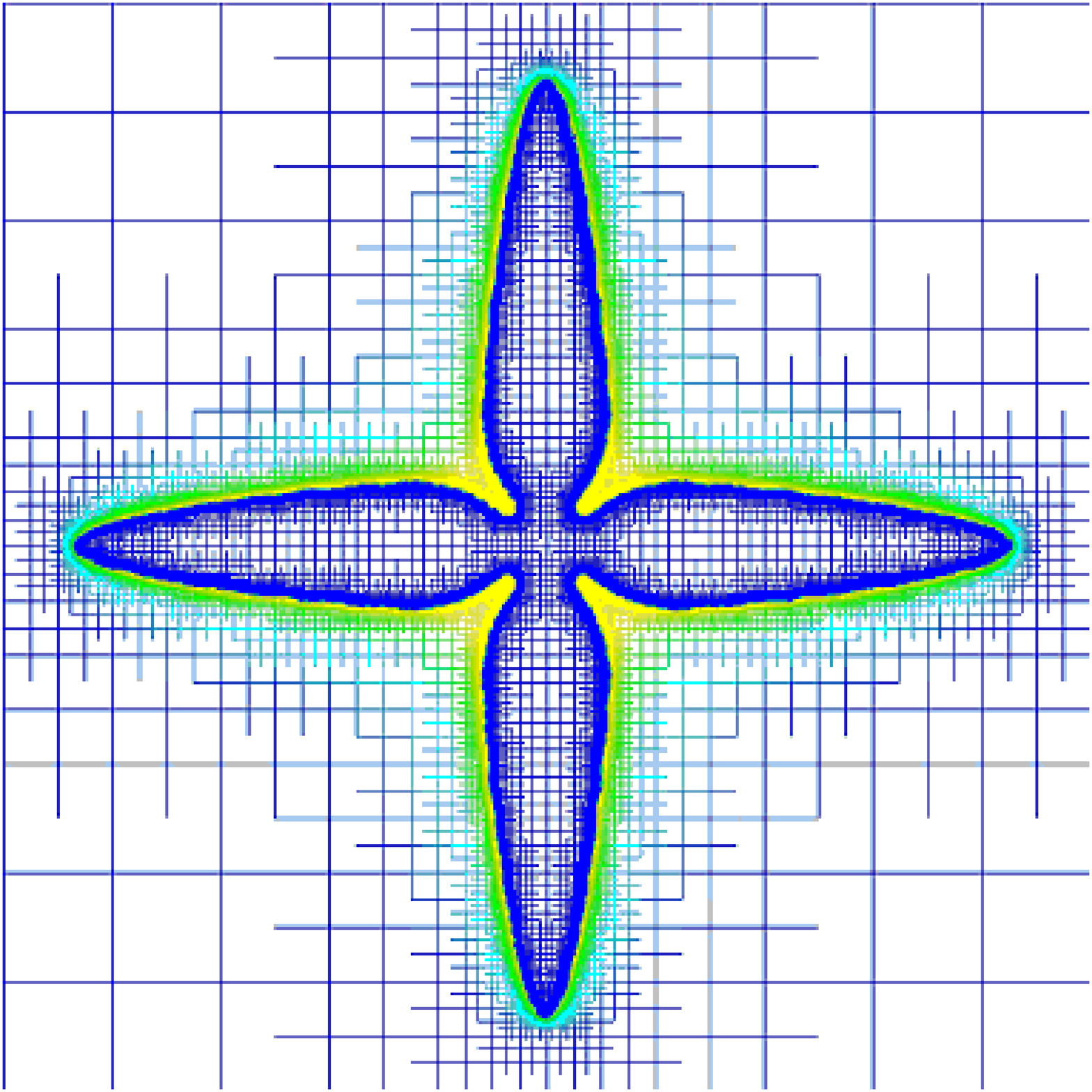}\\
       (c) $C_o=0.05$,$\beta = 400$ and $\epsilon_4 = 0.05$
   \end{tabular}
    \caption[]{\emph{(Colour Online) Morphologies of growth precipitates for a variation in the surface energy anisotropy by modification of $\epsilon_4$.
    The composition field is plotted on the adaptive mesh, blue being low concentration and yellow is a higher concentration.
    The total system size is $6400W$, plotted is an area of $2000W$ on one side. } }\label{MorphTrans3}
\end{figure}

The surface energy anisotropy is entered into the model using the
simple form for 4-fold surface energy $\gamma = \gamma_o(1 -
\epsilon_4 cos(4 \theta))$.  The effect of surface energy anisotropy
is examined by varying $\epsilon_4$ and holding both $C_o$ (or equivalently
the super saturation $\Omega=0.555$) and the elastic anisotropy ($\beta=400$)
constant. Figure \ref{MorphTrans3} (a-c) shows the effect of
increasing the strength of the surface anisotropy under  these
conditions. From top to bottom in Figure \ref{MorphTrans3}
the values of the $\epsilon_4$ used are $\epsilon_4 =
0.01,0.03,0.05$ respectively. Analogously with the effect shown in section
\ref{Morph:beta}, increasing the strength
of $\epsilon_4$ causes the morphology to transform from a
preferential growth along the [11] direction ( Figure
\ref{MorphTrans3} (a) ) to that of the [10] direction ( Figure
\ref{MorphTrans3} (c) )with a transition region where the
precipitate structure is isotropic ( Figure \ref{MorphTrans3} (b) ).

\subsection{Supersaturation($\Omega$)}

\begin{figure}[h]
    \centering
    \begin{tabular}{ccc}
        \includegraphics[height=0.2\textheight]{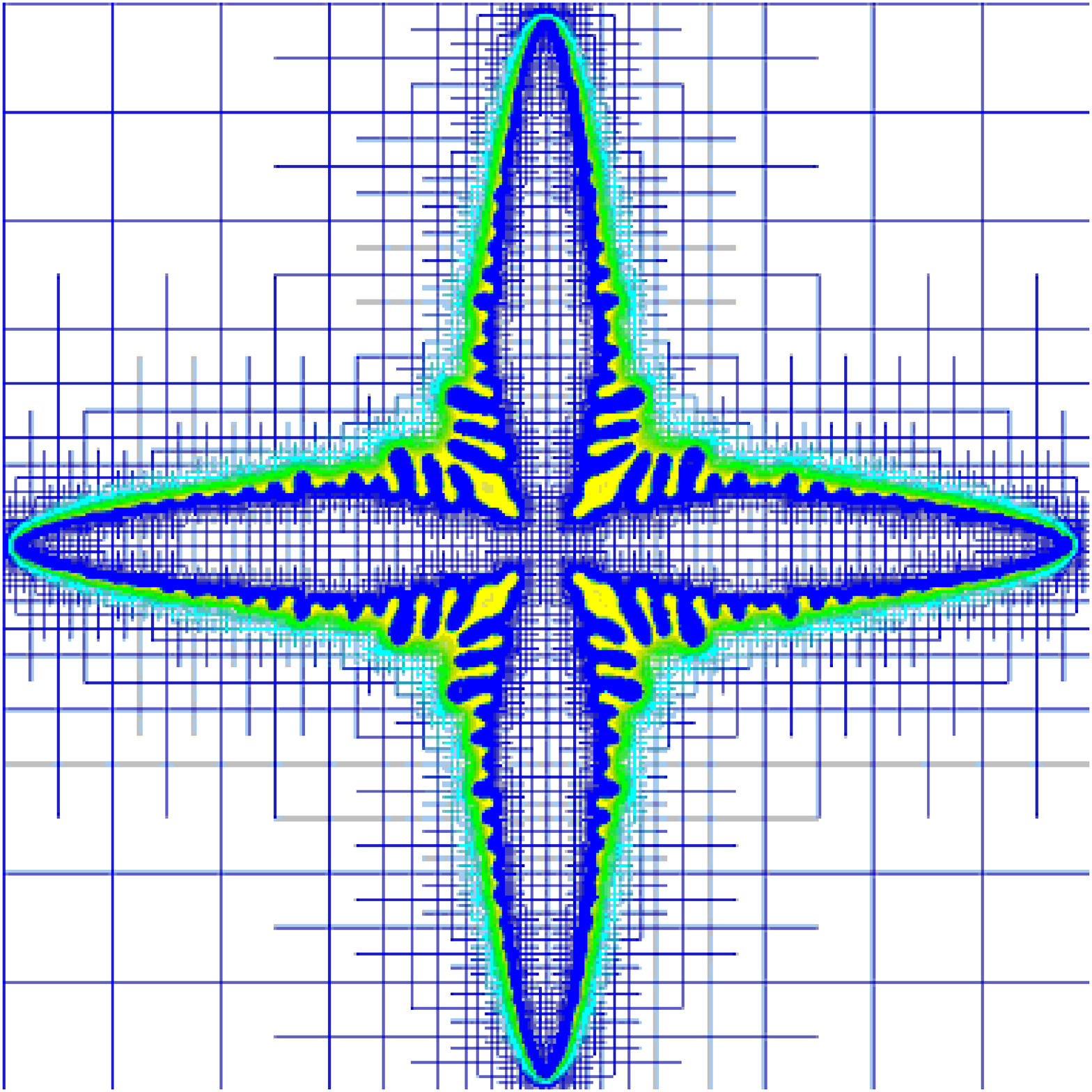}\\
(a) $C_o=0.04$,$\beta = 200$ and $\epsilon_4 = 0.03$\\
        \includegraphics[height=0.2\textheight]{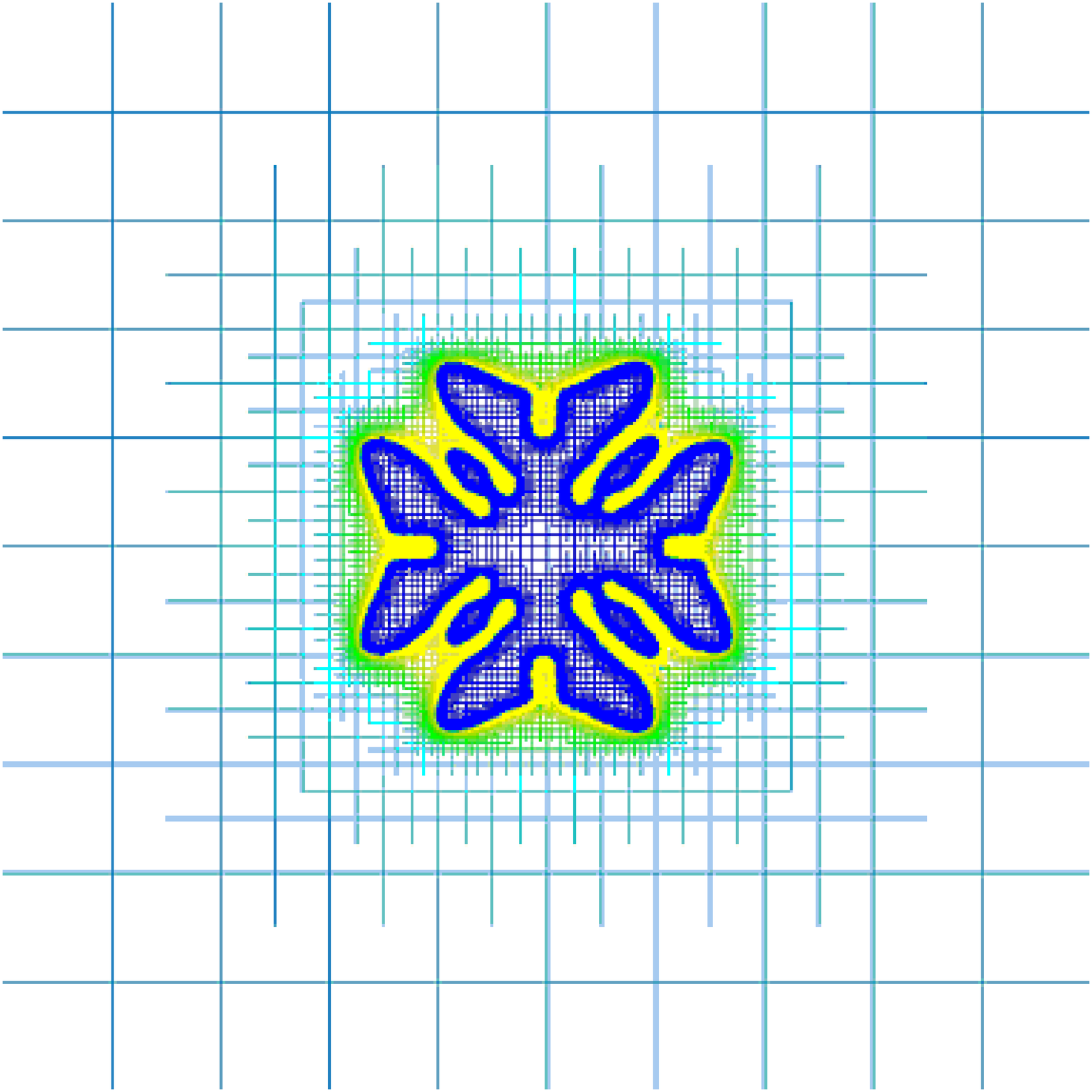}\\
(b) $C_o=0.06$,$\beta = 200$ and $\epsilon_4 = 0.03$\\
        \includegraphics[height=0.2\textheight]{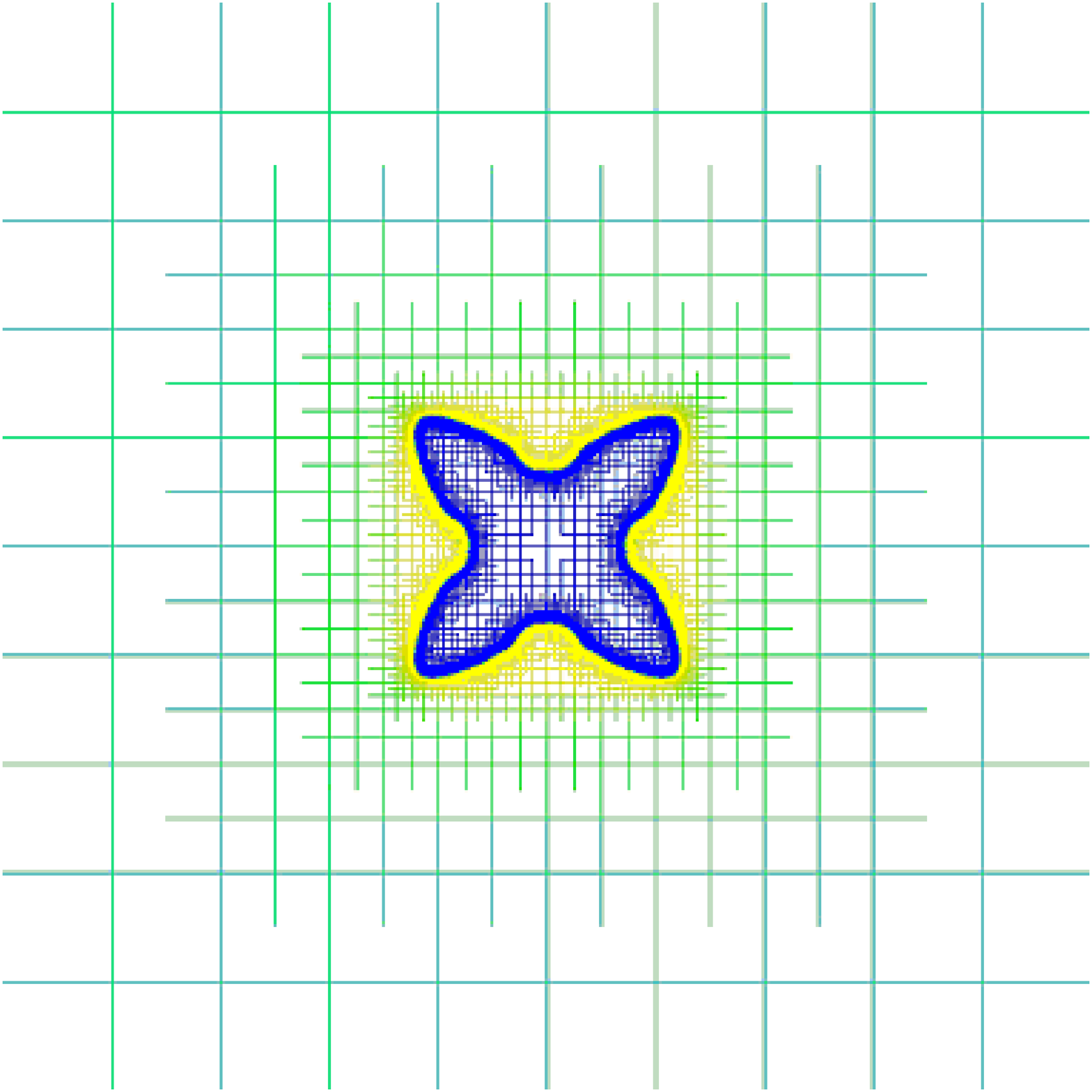}\\
        c) $C_o=0.08$,$\beta = 200$ and $\epsilon_4 = 0.03$
    \end{tabular}
    \caption[]{\emph{(Colour Online) Morphologies of growth precipitates for a variation in the supersaturation with both anisotropic strengths, $\epsilon_4$ and $\beta$, held constant.
    The composition field is plotted on the adaptive mesh, blue being low concentration and yellow is a higher concentration.
    The total system size is $6400W$, plotted is an area of $2000W$ on one side.} }\label{MorphTrans1}
\end{figure}

The supersaturation is varied by changing the initial (i.e. average) alloy
composition $C_o$. Figure \ref{MorphTrans1} (a-c) shows the effect
of decreasing the supersaturation while holding the strength of the
surface anisotropy and elastic anisotropy constant. From top to
bottom the values of the initial alloy composition used are $C_o =
0.04,0.06,0.08$ respectively.  For large super saturations (Figure
\ref{MorphTrans1} (a)) the growth direction is dominated along
directions preferred by the surface energy, i.e. the [10]
directions.  As the supersaturation is decreased a transition from
the [10] growth direction to the [11] direction is observed ( Figure
\ref{MorphTrans1} (b) and (c) ).

\section{Characterization of the Morphological Transition\label{Characterization}}

In this section we discuss a technique by which
the dominant precipitate growth direction can be predicted.
First, the point
at which this transition occurs is defined and measured by examining
the envelope of the precipitate tips in $R-\theta$ space.
For a given value of the precipitate radius, analysis of the amplitude of the
envelope makes it possible to determine the critical surface
energy anisotropy ($\epsilon_4^C$), for a specified elastic
anisotropy ($\beta$), where a morphological transition from $[10]$ to $[11]$
growth directions occurs.
The elastic strain energy is, however,  proportional to the
precipitate area (in 2D) and the surface energy contribution varies
as the particle perimeter. As such the critical
$\epsilon_4^C$ is also a function of the precipitate size.
The particle size dependency of $\epsilon_4^C$ vs. $\beta$ is then
found by balancing the Gibbs-Thomson
corrections corresponding to surface energy vs. elastic anisotropy.  This critical
radius scale is found to be proportional to the Mullins-Sekerka instability radius
($R^C$).  Finally, a condition relating $\epsilon_4^C$ as a function of $\beta$
and $R^C$ at the transition is proposed.

\subsection{Defining the Transition Point}

\emph{The transition point that characterizes the controlling
mechanism of growth morphology is defined as the point at which all competing
anisotropies exactly cancel.}  Under this condition an isolated
precipitate will grow (ideally) as a circle (a sphere in three
dimensions) until the interface becomes unstable by the
Mullins-Sekerka instability. While in this case the interface will
become unstable, the envelope around the particle will continue to
grow as a spheroid. It is this envelope that allows the point of
transition between anisotropically controlled directions to be characterized.

\begin{figure}[h]
    \centerline{\includegraphics*[height=2in,width=3in]{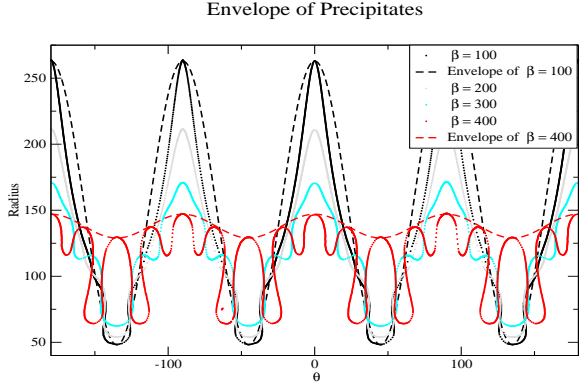}}
    \caption[]{\linespread{0.75} (Colour Online) \em{$R$ - $\theta$ space of 4 precipitates with different  $\beta$ values, $\epsilon_4 = 0.03$ and Supersaturation $= 0.666$.  The dashed lines illustrate the corresponding envelope of the precipitate. As the control values approach a critical point, the envelope amplitude goes to 0. }}
\label{envelope:env}
\end{figure}

The concept of the precipitate envelope is illustrated in Figure
\ref{envelope:env} where the interface for 4 precipitates with
values of $\beta = 100,200,300,400$, $\epsilon_4=0.03$ and $\Omega
=0.666$ in $R-\theta$ space are plotted, where $\Omega =
\frac{C_b^{eq}-C_o}{(1-k)C_b^{eq}}$. The dashed line shows the
envelope surrounding the interface. As the magnitude of $\beta$
approaches the transition point the amplitude of the envelope
decreases, approaching zero.

\subsection{Measurement of the Critical Surface Energy Anisotropy - $\epsilon_4^C$}

The critical surface energy anisotropic coefficient (denoted
$\epsilon_4^C$) is defined as the value of $\epsilon_4$ at a given
supersaturation ($\Omega$) and elastic anisotropy ($\beta$) which
results in an envelope amplitude of zero. $\epsilon_4^C$ is
interpolated from the amplitudes of the precipitate envelopes obtained by
varying $\epsilon_4$ for given values of the supersaturation
$\Omega$ and elastic anisotropy $\beta$.

The envelope amplitude is approximated by measuring the difference
of the total growth distance from the center of the precipitate
along the [10] direction to the growth distance along the [11]
direction. The transition point is the interpolated value for
$\epsilon_4$ such that these amplitudes approach zero. Seven
different supersaturations are considered here, where $\epsilon_4$ is varied
between $0.005$ and $0.05$ and $\beta$ is varied from $100$ to
$400$. The envelope amplitudes are measured at arbitrary times,
chosen in each case, however,  such that the precipitate has outgrown
any initial transients.

\begin{figure}[h]
    \centerline{\includegraphics*[height=2.2in,width=3in]{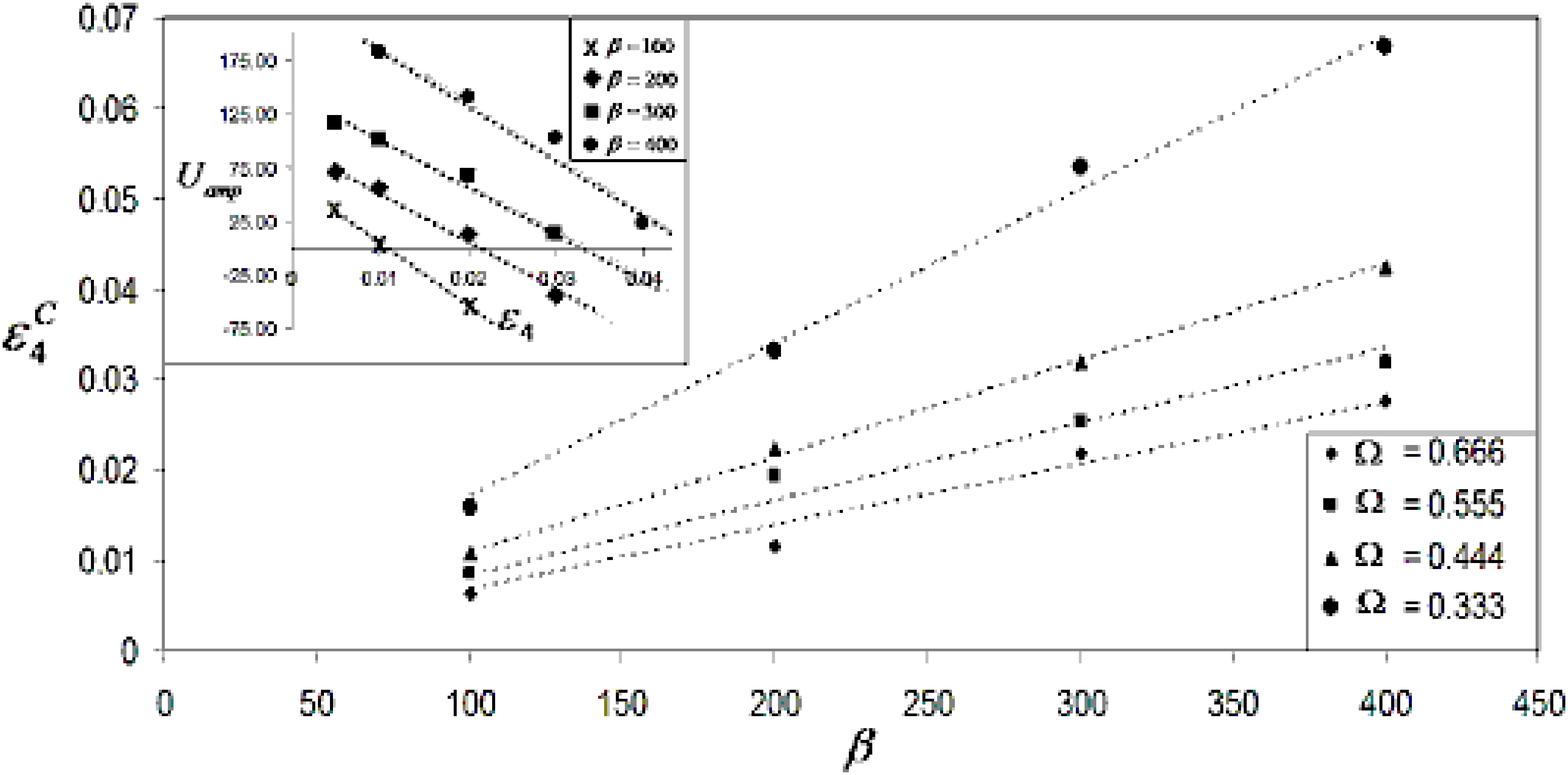}}
    \caption[]{\linespread{0.75}\em{The relationship between $\epsilon_4^C$ and
    $\beta$ is extracted by fitting their plots.  Shown here are the
    $\epsilon_4^C$ vs $\beta$ for $4$ supersaturations, $\Omega = 0.666$,
    $\Omega = 0.555$, $\Omega = 0.444$, $\Omega = 0.333$.  \emph{Inset:} Interpolation
     of $\epsilon_4^C$ is extrapolated by fitting measured values for $U_{amp}\rightarrow 0$
      for $\Omega = 0.444$.}}
\label{critsurf:E4Beta}
\end{figure}

For each value of $\beta$ at each alloy composition $C_o$, the
envelope amplitude is plotted vs $\epsilon_4$. The inset to Figure
\ref{critsurf:E4Beta} illustrates this for an alloy with an average
composition of $C_o = 0.06$, and the deviation from elastic isotropy
is characterized for values of $\beta = 100,200,300,400$. For each
value of $\beta$, the data is fitted linearly and is interpolated to
the transition line to extract the critical surface anisotropic
value $\epsilon_4^C(\beta, \Omega)$.
A linear relationship between $\epsilon_4^C$ and $\beta$ is found
for each of the supersaturations studied. These linear
relationships are plotted in Figure \ref{critsurf:E4Beta} for the
alloy compositions of $C_o = 0.04,0.05,0.06,0.07$. Figure
\ref{critsurf:E4Beta} thus predicts that
\begin{equation}
\epsilon_4^C = A_{C_o}\beta \label{E4Cfit}
\end{equation}
where the fitting parameter $A_{C_o}$ has values of $A_{0.04} = 6.84
\cdot 10^{-5}$,$A_{0.05} = 8.33\cdot 10^{-5}$,$A_{0.06} = 10.68\cdot
10^{-5}$ and $A_{0.07} = 16.98\cdot 10^{-5}$ for the
supersaturations studied.

\subsection{Critical Tip Radii at the Transition Point}

The competing anisotropic effects controlling morphology
cancel when all correction terms that are dependent on the
interface normal angle ($\theta$) in the
interfacial equilibrium composition ( $C_b^{int} = C_b^{eq} - \Delta
C_{capillarity}(1+15 \epsilon_4f(\theta)) - \Delta
C_{elastic}(1-\epsilon_eh(\theta))$ ) exactly cancel. Here $\Delta
C_{capillarity}$ and $\Delta C_{elastic}$ are the isotropic
corrections to interfacial equilibrium composition and $\epsilon_4$
and $\epsilon_e$ represent the strength of the surface energy and
elastic energy anisotropies. Assuming a linear fourier expansion
with a 4-fold symmetry in both $f(\theta)$ and $h(\theta)$ (ie.
$cos(4\theta)$) the terms in the interface solute correction can be grouped by
order of the fourier expansion. \ben C_b^{int} = C_b^{eq} - (\Delta
C_{capillarity} + \Delta C_{elastic}) \nonumber \\
- \left(\epsilon_{surf} - \epsilon_{ele}\right)cos(4\theta)
\label{Morph:Cbint}\een where $\epsilon_{surf} = 15 \epsilon_4
\Delta C_{capillarity}\kappa$ and $\epsilon_{ele} = \epsilon_e
\Delta C_{elastic}$ are the relative anisotropic strengths of the
surface energy correction and the elastic energy correction
respectively. The factor of $15$ in the capillarity term comes from
the stiffness of the capillarity, $d_o(\vec{n}) = d_o(1+15\epsilon_4
cos(4 \theta))$. The elastic anisotropy strength, $\epsilon_e$,
is linked to the strength of the elastic anisotropy through $\beta$ and
is derived  below (see Equation \ref{EE:epse}).

For an isotropic morphology to emerge the coefficient of $\cos(\theta)$ in Equation
\ref{Morph:Cbint} is required to vanish, ie. \ben \epsilon_{surf} -
\epsilon_{ele} = 0\label{Morph:coszero}\een In Equation \ref{Morph:Cbint} the capillary term ($\Delta
C_{capillary}$) contains a curvature correction, while the elastic
term ($\Delta C_{elastic}$) does not. When these terms balance each
other a curvature $\kappa$ is selected, which will be associated with a
critical radius $R_{trans}^C$ (ie. $\kappa_c=1/R_{trans}^C$). This is determined next.

The capillarity correction is $\Delta C_{capillarity} =
(1-k)d_oC_b^{eq}$, which is used to calculate $\epsilon_{surf}$ as
 \ben \epsilon_{surf} =
15C_b^o(1-k)d_o\epsilon_4\kappa\label{Morph:Esurf}\een  The total
elastic correction for cubic coefficients is defined by Equation
\ref{Ele:Cb} as $ \Delta C_{elastic} = \frac{1}{2}\frac{Z_1}{1-k} =
\frac{1}{4}\frac{C_{11}+C_{12}}{1-k}\epsilon^{*2}(1-\frac{\epsilon_{xx}+\epsilon_{yy}}{\epsilon^*})
$. In the absence of elastic anisotropy the strain trace
($\epsilon_{xx}+\epsilon_{yy}$) is zero and therefore \ben \Delta
C_{elastic} = \frac{1}{4}\frac{C_{11}+C_{12}}{1-k}\epsilon^{*2} \een
$\epsilon_{ele}$ is calculated by consideration of the anisotropy in
the strain field by substituting Equation \ref{EE:Exxyy} for
$\epsilon_{xx}+\epsilon_{yy}$, giving \ben
1-\frac{\epsilon_{xx}+\epsilon_{yy}}{\epsilon^*} =
 1-\epsilon_e
cos(4\theta)\label{EE:epse}\een where $\epsilon_e =
\frac{U_{amp}}{\epsilon^*}$ and $U_{amp}$ is defined by Equation
\ref{AnisElast:Uamp}. This results in $\epsilon_{ele}$ becoming
\ben\epsilon_{ele}=\frac{1}{8(1-k)}\frac{C_{11}+C_{12}}{C_{11}+\beta}\beta\epsilon^{*2}\label{Morph:Eele}\een
Substituting Equations \ref{Morph:Esurf} and \ref{Morph:Eele} into
Equation \ref{Morph:coszero}, the critical radius of
curvature required to maintain isotropic conditions is given by,
\begin{equation}
R_{trans}^C = 120(1-k)^2C_o^ld_o\frac{
(C_{11}+\beta)}{\epsilon^{*2}(C_{11}+C_{12})}\frac{\epsilon_4^C}{\beta}
\label{Morph:RCE4Beta}
\end{equation}
A fit to a selected critical radii is attained by substituting the
fitted equation for the critical surface anisotropy coefficient
(Equation \ref{E4Cfit}) and by choosing a reference point of $\beta
= 0$.  This results in a relationship for the magnitude of the
$\beta = 0$ transition radius that depends on concentration through
$A_{C_o}$ given as,
\begin{equation}
R_{trans} = \frac{120
(1-k)^2(C_{11})C_o^ld_o}{(C_{11}+C_{12})\epsilon^{*2}}A_{C_o}
\label{Morph:ERfit}
\end{equation}
The next subsection discusses a separate method by which this radius is estimated without the need for
measured values of $A_{C_o}$.

\subsection{Calculation of the Transition Tip Radius using Linear Stability Approximation}

In the previous section a selected precipitate radius is derived
based on the interpolated values for the critical surface energy
anisotropy. However, this method relates $\epsilon_4^C$ to $\beta$
only once the value of $A_{C_o}$ is measured.  A method to
approximate the selected radius by theoretical consideration of the
Mullins-Sekerka linear stability analysis on an isotropic particle
is now shown.

Mullins and Sekerka in 1963 \cite{Mul63,Lan80} predicted a critical
particle size $R^*_{k=2} = 11R^*$ ($R^*=\frac{2d_o}{\Omega}$, $k$ is
the instability mode and $\Omega$ is the supersaturation) after
which the particle interface becomes unstable.   A particle at the
transition point in our study can be considered to behave similarly
to an isotropic particle in a supercooled matrix and the tip's
radius of curvature is assumed to be proportional to this critical
radius. Here the supersaturation of the precipitate under elastic
strain is modified by the elasticity according to $\Omega_{el}
=\frac{C_b-C_o}{(1-k)C_b}$, where $C_b$ is the equilibrium interface
composition modified due to elasticity  by Equation \ref{Ele:Cb}.
This supersaturation is used in the linear stability analysis result
to predict a minimum critical radius of,
\begin{equation}
R_{MS} = 22\frac{d_o}{\Omega_{el}} \label{Eq25}
\end{equation}

\begin{figure}[h]
    \centerline{\includegraphics*[height=2.5in,width=3in]{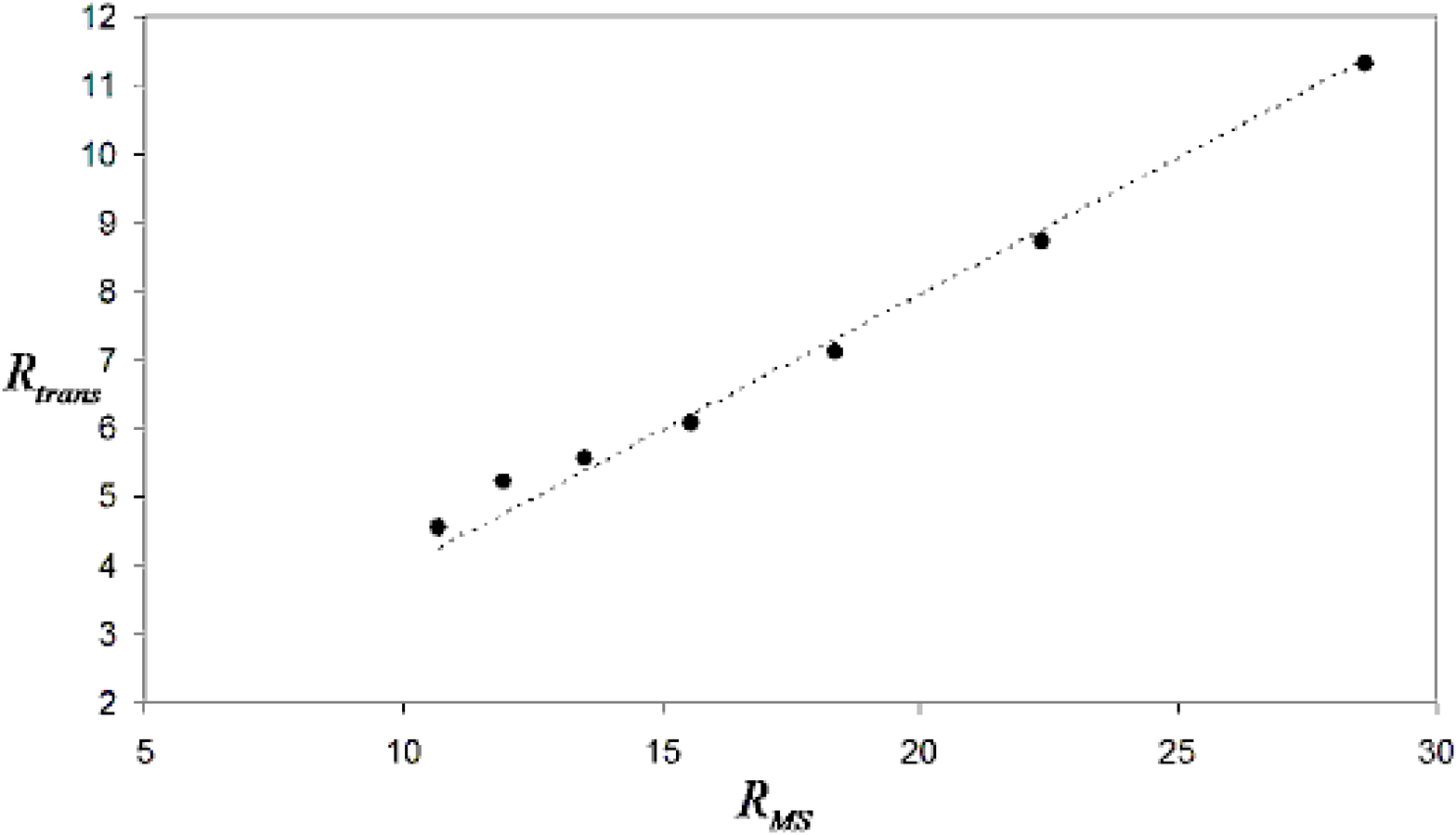}}
    \caption[]{\linespread{0.75}\em{The fitted critical tip radius
    for isotropic growth, from equation \ref{Morph:ERfit},
     vs the onset Mullins-Sekerka wavelength for seven different supersaturations.}}
\label{Morph:RfitRms}
\end{figure}

A comparison of this instability radius with the fitted radius of
equation \ref{Morph:ERfit} shows excellent linear agreement as shown
in figure \ref{Morph:RfitRms}.
Equation \ref{Morph:ERfit} is calculated such that $C_{11} + \beta
\rightarrow C_{11}$ in Equation \ref{Morph:RCE4Beta}.  The
 $\beta$  dependence of $R_{trans}^C$ is introduced into Equation~\ref{Eq25} through its
relationship to $\epsilon_e$ by the term $\frac{1}{C_{11}+\beta}$.
The final result for the critical selected radius as a function of
$\beta$ and $\Omega_{el}$ is
\begin{equation}
R^{MS}_{trans} =
\frac{44}{5}\frac{d_o}{\Omega_{el}}(1+\frac{\beta}{C_{11}})
\label{LS:RC}
\end{equation}

\subsection{Calculation of the Critical Transition Point}

With the scale of the critical tip radius determined by linear stability theory as
a function of supersaturation, the required measurement of the relationship
between $\epsilon_4^C$ and $\beta$ can be eliminated in Equation~\ref{Morph:RCE4Beta}.
This is done by substituting the linear
stability prediction of the critical radius (Equation \ref{LS:RC})
into the equation for the selected transition radius (Equation \ref{Morph:RCE4Beta}). The resulting
 relationship is solved for the critical surface energy
anisotropy
 pre-factor ($\epsilon_4^C$). This relationship is
\begin{equation}
\epsilon_4^C =
\frac{11}{150}\frac{C_{11}+C_{12}}{(1-k)^2(C_{11}+\beta)C_o^l\Omega_{el}}\epsilon^{*2}\beta
(1+\frac{\beta}{C_{11}})\label{Morph:E4Ccrit}
\end{equation}

\begin{figure}[h]
    \centerline{\includegraphics*[height=2.5in,width=3in]{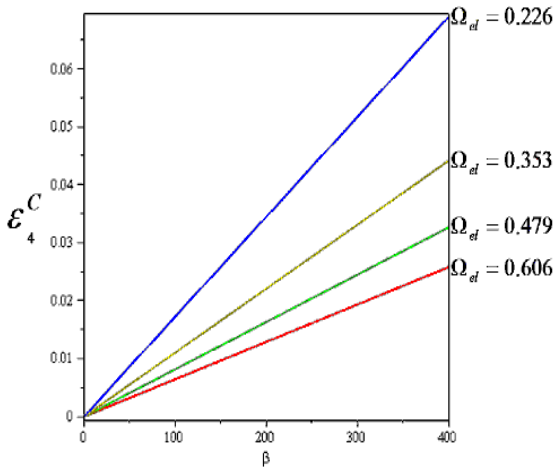}}
    \caption[]{\linespread{0.75}\em{The critical transition points from equation
    \ref{Morph:E4Ccrit}. Above the line the precipitates prefer to
    grow in directions that minimize the surface energy [10], and below
    the directions which minimize the elastic energy [11]. Transition lines for different $\Omega_{el}$ are plotted. Colour Online.}}
\label{TransLines4}
\end{figure}

\begin{figure}[h]
    \centerline{\includegraphics*[height=3in,width=3in]{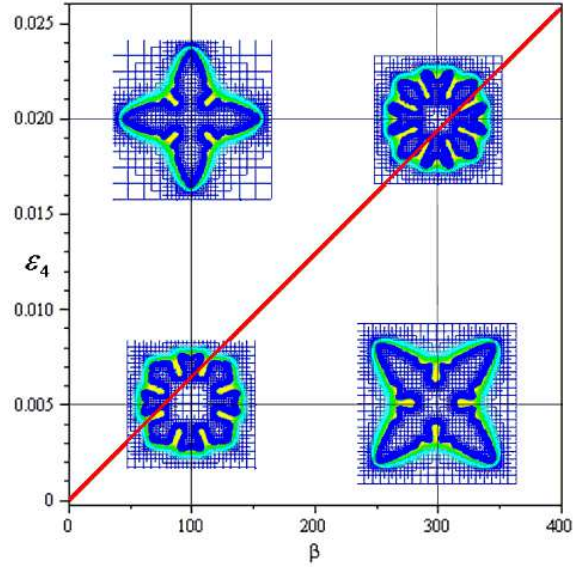}}
    \caption[]{\linespread{0.75}\em{The critical transition line from equation
    \ref{Morph:E4Ccrit}. Above the line the precipitates prefer to
    grow in directions that minimize the surface energy [10], and below
    the directions which minimize the elastic energy [11]. The curve
     is plotted vs beta for $C_o = 0.04$($\Omega_{el} = 0.606$).
     Overplotted are 4 simulations corresponding
     to different values of  $\epsilon_4$ and $\beta$.  The plotted
     line represents
     the value of $\epsilon_4^C$, the transition point where
     anisotropies cancel out. Colour Online.}} \label{TransLines1}
\end{figure}

Equation~\ref{Morph:E4Ccrit} defines a morphological transition line
as a function of supersaturation ($\Omega$), elastic anisotropy
($\beta$) and the anisotropy of the capillarity ($\epsilon_4$).  The
transition lines for $\Omega_{el} = 0.606,0.479,0.353,0.226$ are
plotted in figure \ref{TransLines4}.  Precipitates grown above the
transition line will grow in the [10] directions while growth for
conditions below the line will grow along [11] directions. Some
morphologies are overplotted above and below the transition line for
$\Omega = 0.606$ in figure \ref{TransLines1} to further illustrate
the utility of Equation~\ref{Morph:E4Ccrit}.

\section{Summary}

We have introduced a phase-field model for the study of the
morphological development of elastically stressed solid state
precipitates.  We considered particles that have coherent interfaces and are under
elastic self stress by a lattice mismatch eigenstrain.  We used a new
finite difference based adaptive mesh refinement algorithm  to solve the phase-field
and strain relaxation equations thereby allowing for very rapid solution times.
By investigating the effects of supersaturation, elastic anisotropy in the elastic tensor and
anisotropy in the capillarity we developed a scaling relationship to
predict which anisotropy will be dominant in the morphological evolution
of the precipitate.  It is  interesting note is the effect of
supersaturation on the selected precipitate morphology and growth
directions.

We would like to thank the National Science and Engineering Research Council of Canada
(NSERC) for financial support and SHARC-NET for
supercomputing support.

\appendix

\section{Cubic Elastic Free Energy Coefficients\label{Ele:Cubic}}

In the generalized elastic portion of the phase field free energy,
as described by equation \ref{EE:felNV} $ f_{el} ={\it Z_3}  \left(
g_3(\phi) \right) ^{3}+{\it Z_2}  \left( g_3
 \left( \phi \right)  \right) ^{2}+{\it Z_1} g_3(\phi) +{ \it Z_0}
$,
 several unknown terms $Z_3$, $Z_2$,
$Z_1$ and $Z_0$ are introduced. These coeffiecients are dependent on
the particular values of the elastic modulus tensor in either of the
precipitate or matrix phases. Presented here are the explicit forms
for these functions for two sided cubic modula and a hydrostatic
elastic eigenstrain of the form,
\[ \epsilon^*_{ij} = \left| \begin{array}{ccc}
\epsilon^* & 0 \\
0 & \epsilon^*  \end{array} \right|.\]  The zeroth order
component($Z_0$) has no dependence on the phase at all and is
calculated to be \ben Z_0 = \frac{1}{4}(C1_{11} +
C2_{11})(\epsilon_{xx}-\frac{\epsilon^*}{2})^2 \nonumber \\
+\frac{1}{4}(C1_{11}+C2_{11})(\epsilon_{yy}-\frac{\epsilon^*}{2})^2 \nonumber \\
+\frac{1}{2}(C1_{12}+C2_{12})(\epsilon_{xx}-\frac{\epsilon^*}{2})(\epsilon_{yy}-\frac{\epsilon^*}{2}) \nonumber \\
+(C1_{44}+C2_{44})\epsilon_{xy}^2 \een This pre-factor has no
dependence of phase (nor concentration) and therefore it does not
appear in either the phase mobility equation (equation \ref{EE:PDM})
or the chemical diffusion equation (Equation \ref{EE:chemdiff})
since the growth kinetics are dependent on differences of energy. It
does however appear in the static elasticity equation (Equation
\ref{Ele:StEle}).

The first order component is the most prominent term in the model
equations and is calculated to be \ben
 Z_1 = \frac{1}{8}(3(C1_{11}+C1_{12}) + C2_{11} +
 C2_{12})\epsilon^{*2} \nonumber \\
 - \frac{1}{2}(C1_{11}+C1_{12})(\epsilon_{xx}+\epsilon_{yy})\epsilon^*\nonumber \\
 +\frac{1}{4}(C1_{11} -C2_{11})(\epsilon_{xx}^2+\epsilon_{yy}^2)\nonumber \\
 + \frac{1}{2}(C1_{12} - C2_{12})\epsilon_{xx}\epsilon_{yy}\nonumber \\
 +(C1_{44} - C2_{44})\epsilon_{xy}^2
\een

The second order coefficient to the elastic energy in terms of the
phase is calculated to be \ben Z_2 = \frac{1}{8}(3(C1_{11} +C1_{12})
-
(C2_{11}+C2_{12}))\epsilon^{*2} \nonumber \\
+\frac{1}{4}(C2_{11}+C2_{12}-C1_{11}-C1_{12})(\epsilon_{xx} +
\epsilon_{yy})\epsilon^* \een and in the presence of equal elastic
coefficients this term becomes a constant.

The third order component is calculated to be
\begin{equation}
Z_3 = \frac{1}{8}(C1_{11} - C2_{11} + C1_{12} -
C2_{12})\epsilon^{*2}
\end{equation}
and has no dependence on the dynamic strain field. In the presence
of equal elastic coefficients this term vanishes completely.

\bibliography{biblio}

\end{document}